\documentclass[aps,prb,twocolumn,superscriptaddress,nobibnotes,floatfix,longbibliography]{revtex4-2}

\usepackage[utf8]{inputenc}
\usepackage[T1]{fontenc}

\usepackage[version=4]{mhchem} 

\usepackage{color}
\usepackage{epstopdf}
\usepackage{graphicx}
\usepackage{xcolor}
\usepackage{comment}
\usepackage{bm}
\usepackage{amsmath, amsthm, amssymb, amsfonts}
\usepackage{mathtools}
\usepackage{multirow}
\usepackage{array}
\usepackage{dblfloatfix}

\usepackage[colorlinks=true,
            citecolor=blue,
            linkcolor=blue,
            urlcolor=blue,
            bookmarksnumbered=true,
            bookmarks=true,
            bookmarksopen=true,
            unicode=false]{hyperref}

\usepackage{float}

\begin{document}

\title{Cotunneling theory and multiplet excitations: emergence of asymmetric line shape in inelastic scanning tunneling spectroscopy of correlated molecules on surfaces}

\author{Marco Lozano}
\affiliation{Institute of Physics, Czech Academy of Sciences, Prague 16200, Czech Republic}

\author{Manish Kumar}
\affiliation{Institute of Physics, Czech Academy of Sciences, Prague 16200, Czech Republic}
\affiliation{Department of Condensed Matter Physics, Faculty of Mathematics and Physics, Charles University, CZ-12116 Prague 2, Czech Republic}

\author{Pavel Jel\'inek}
\thanks{Corresponding author: \href{mailto:soler@fzu.cz}{jelinekp@fzu.cz}}
\affiliation{Institute of Physics, Czech Academy of Sciences, Prague 16200, Czech Republic}
\affiliation{Czech Advanced Technology and Research Institute (CATRIN), Palack\'y University Olomouc, 779 00 Olomouc, Czech Republic}

\author{Diego Soler-Polo}
\thanks{Corresponding author: \href{mailto:soler@fzu.cz}{soler@fzu.cz}}
\affiliation{Institute of Physics, Czech Academy of Sciences, Prague 16200, Czech Republic}


\begin{abstract}

Recent advances in on-surface chemistry, combined with scanning probe microscopy, have enabled the synthesis of correlated molecules on surfaces and the characterization of their chemical and electronic properties with unprecedented spatial resolution. Low-energy magnetic excitations of individual molecules are frequently investigated by scanning tunneling spectroscopy (STS) and often appear as symmetric step-like features in the differential conductance as a function of bias voltage. The interpretation of such steps is well established within cotunneling theory and effective model Hamiltonians (e.g., Hubbard- and spin-based models). Here, we extend the cotunneling formalism to general multireference systems. We show that multireference character, together with orbital-dependent and strongly asymmetric tip/substrate couplings, can produce pronounced asymmetric line shapes in inelastic STS. These results provide an alternative microscopic mechanism for the asymmetric peaks and dips near the Fermi level frequently observed in STS experiments.
\end{abstract}

\maketitle
\section{Introduction}

Scanning tunneling spectroscopy (STS) has become an indispensable tool to probe electronic, vibrational and magnetic excitations in individual atoms and molecules on surfaces \cite{feenstra1994scanning, stipe1998single,gauyacq2012excitation,li1998kondo,Madhavan1998}. In particular, the appearance of low-bias features such as steps or zero energy peaks in the differential conductance ($dI/dV$) has been frequently associated with many-body excitations at the nanoscale \cite{schulz2015many, neel2007controlled,oberg2014control}. 
Most studies have associated symmetric step-like features in $dI/dV$ with spin excitations mediated by magnetic anisotropy \cite{fernandez2009theory, ormaza2017controlled,ternes2015,Heinrich2004,Loth2010} or transitions between the ground and first excited states \cite{Tenorio2024,Zuzak2024}. On the other hand, sharp zero-bias peaks are typically attributed to Kondo screening \cite{warner2016sub,Vegliante2025}. 

However, often STS spectra reveal the presence of asymmetric low-bias anomalies, where the unambiguous identification of their physical origin is still challenging. 
For example, the origin of zero bias anomalies with characteristic dip feature observed experimentally on magnetic adatoms on a metal surface \cite{Madhavan1998,Knorr2002} are still the subject of intensive debate. Originally, the presence of characteristic dip at zero bias was associated with a Fano resonance \cite{Fano1961} due to interference between tunnelling channels through the Kondo state and surrounding electronic states of surface or Kondo impurity \cite{jsghy2000,Merino2004,Tacca2021,Fernndez2021}. Recently, alternative explanation of this zero bias anomaly was proposed within the framework of so called spinaron, many-body localized state, originating from spin excitations on magnetic impurity and their interaction with conduction electrons \cite{Bouaziz2020,Friedrich2023}.

Another complexity comes from the fact that the electronic structure of many magnetic atoms or open-shell molecules \cite{Kumar2025,Tenorio2024,Zuzak2024,Vegliante2025} have strong multireference character, i.e. the ground and excited states cannot be simply represented by a single Slater determinant. Therefore, the proper theoretical description of STS spectra requires the use of a multi-reference states including virtual transition charge states within the cotunneling formalism \cite{delgado2011cotunneling, Ternes2008}. Here the multireference states consists of several Slater determinants with different weights represented in the basis of one-electron atomic/molecular orbitals. In principle, each one-electron orbital may have different spatial extension, which can provide its highly asymmetric coupling to electronic states of the tip and/or sample. This raises fundamental questions about how the non-homogeneous coupling between the tip and one-electron molecular states influences the character of the IETS line shape in multireference systems.

In this work, we present a cotunneling theory for multireference molecular systems weakly adsorbed on the Au(111) surface. We investigate theoretically the character of spin excitations line shapes in the multireference molecular systems, where the coupling between tip/sample and molecular orbitals is strongly asymmetric. We demonstrate that the disproportional coupling in multireference systems can give rise to a strongly asymmetric STS line shape, in sharp contrast to the highly symmetric spectral features associated with spin excitation originated from the magnetic anisotropy due to spin–orbit coupling.

\section{Theoretical modeling}

\begin{figure}[h]
    \centering
    \includegraphics[width=1\linewidth]{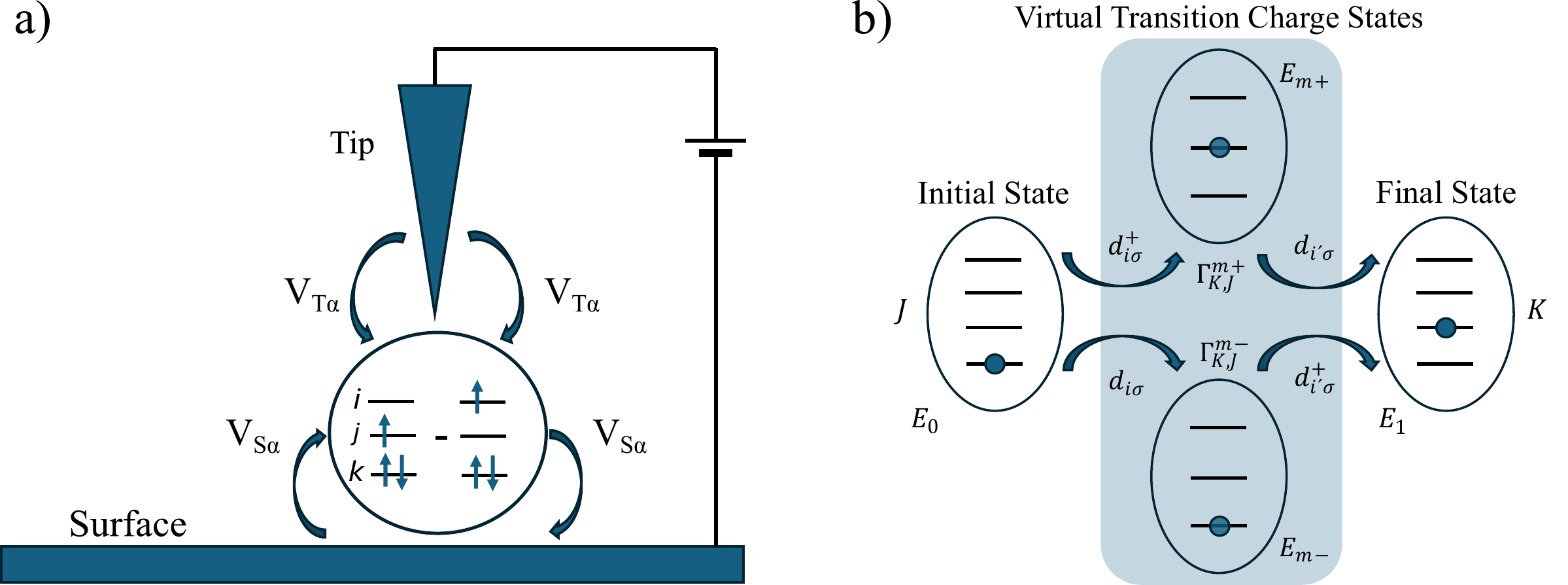}
    \caption{Cotunneling scheme for a molecular quantum dot. a) The representation of the ground state as a combination of configurations with different orbital occupations (\emph{i,j,k}). The Tip and Surface couple to the orbital sites ($\alpha=i,j,k$). b) Representation of an excitation mediated by virtual charge transitions. The blue dot accounts for the state in which the system is found in the different $N$ and $N \pm1$ situations.}
    \label{fig:scheme model}
\end{figure}

We consider a molecule weakly coupled to a metallic surface, which is inspected with a STM tip in the tunneling regime, see FIG. \ref{fig:scheme model} a). Under such conditions and in the absence of charge transfer between the molecule and the surface, the electronic structure of the molecule is well approximated by neutral state in the gas phase. The theoretical model consists of a quantum dot Hamiltonian representing the molecule, which is connected to metallic leads,  Tip (T) and Substrate (S), respectively. The total Hamiltonian can be written as:
\begin{align}
\hat{H} &=
\underbrace{\sum_{ij\sigma} t_{ij} d_{i\sigma}^\dagger d_{j\sigma}
+ \frac{1}{2} \sum_{ijkl \atop \sigma\sigma'}
U_{ijkl} d_{i\sigma}^\dagger d_{j\sigma'}^\dagger
d_{k\sigma'} d_{l\sigma}}_{\hat{H}_{\text{dot}}}
\nonumber \\
&\quad + \underbrace{\sum_{k\sigma} \epsilon_{T k}
c_{T k\sigma}^\dagger c_{T k\sigma}}_{\hat{H}_{\text{T}}}
+ \underbrace{\sum_{k\sigma} \epsilon_{S k}
c_{S k\sigma}^\dagger c_{S k\sigma}}_{\hat{H}_{\text{S}}}
\nonumber \\
&\quad + \underbrace{\sum_{\alpha = T,S} \sum_{ik\sigma}
\left( V_{\alpha ik} c_{\alpha k\sigma}^\dagger d_{i\sigma}
+ \text{H.c.} \right)}_{\hat{H}_{\text{tun}}}.
\label{eq:basic_model}
\end{align}

where $d_{i\sigma}^\dagger$ ($d_{i\sigma}$) creates (annihilates) an electron with spin $\sigma$ at the molecular level $i$ on the quantum dot, $t_{ij}$ are the hopping terms, and $U_{ijkl}$ are the elements of the Coulomb matrix.

\begin{equation}
    H =  H_{\text{dot}} + H_{\text{T}} + H_{\text{S}}  +H_{\text{t}}
    \label{eq: basic model}
\end{equation}
The quantum dot is described by a general many-body Hamiltonian:
\begin{equation}
    H_{\text{dot}} = \sum_{ij\sigma} t_{ij} d_{i\sigma}^\dagger d_{j\sigma} + \frac{1}{2} \sum_{ijkl\atop \sigma\sigma'} U_{ijkl} d_{i\sigma}^\dagger d_{j\sigma'}^\dagger d_{k\sigma'} d_{l\sigma}.
    \label{eq:dot}
\end{equation}
where $d_{i\sigma}^\dagger$ ($d_{i\sigma}$) creates (annihilates) an electron with spin $\sigma$ at level $i$ on the dot, $t_{ij}$ are the single-particle hopping terms, and $U_{ijkl}$ are the elements of the Coulomb matrix. Note that the solution of such a Hamiltonian is a set of multiplets whose wave-functions are represented as expansion of Slater determinants in the orbital basis sites used to write the Hamiltonian.

The metallic leads are given by non-interacting electron reservoirs $H_{\alpha} = \sum_{k\sigma} \epsilon_{\alpha k} c_{\alpha k\sigma}^\dagger c_{\alpha k\sigma}$, where $c_{\alpha k\sigma}^\dagger$ ($c_{\alpha k\sigma}$) creates (annihilates) an electron with momentum $k$ and spin $\sigma$ in lead $\alpha$ ($\alpha = \text{T,S}$). The last term in eq. \eqref{eq: basic model} is a dot-lead coupling in the form of $H_{\text{t}} = \sum_{\alpha=T,S} \sum_{ik\sigma} \left( V_{\alpha ik} c_{\alpha k\sigma}^\dagger d_{i\sigma} + \text{H.c.} \right)$, with $V_{\alpha ik}$ being the tunneling amplitude between lead $\alpha$ and level $i$ on the dot. A scheme describing the cotunneling is shown in FIG. \ref{fig:scheme model} b), where an excitation in the neutral system is achieved through the virtual charging processes.
\vspace{0.1cm}

Given the model in eq. \eqref{eq: basic model}, we resort to the cotunneling formalism \cite{delgado2011cotunneling,begemann2010inelastic}, which describes the contribution to inelastic current from second-order processes such as spin excitations. In this framework, we first solve by exact diagonalization the Hamiltonian $H_\text{dot}$ 
and then treat perturbatively the tunneling terms $H_\text{tun}$ in order to get the expression for the current $I_T$:

\begin{equation}
\label{eq: current cotunneling}
    I_T = e \sum_{JK} P_K(V) \left( W^{S\rightarrow T}_{J\rightarrow K} - W^{T\rightarrow S}_{J\rightarrow K} \right)
\end{equation}
where $P_K(V)$ is the population of states for the neutral system and $W^{S\rightarrow T}_{J\rightarrow K}$ are the transition rates between many-body state $E_J$ and $E_K$ given by
\begin{align}
W_{J\rightarrow K}^{\alpha \rightarrow \alpha'} &=
\dfrac{2 \pi}{\hbar} \int d \epsilon \,
\rho_\alpha(\epsilon) \rho_{\alpha'}(\epsilon + \Delta_{KJ})
\nonumber \\
&\times \sum_{i,i'} \sum_{m+, m-} \sum_{\sigma, \sigma'}
\Big|
\dfrac{V_{\alpha i k} V_{\alpha' i' k}}
{E_{m+} - E_0 - \epsilon_{\alpha}}
\Gamma_{K,J}^{m+}(ii',\sigma \sigma')
\nonumber \\
&\quad -
\dfrac{V_{\alpha i k} V_{\alpha' i' k}}
{E_{m-} - E_0 - \epsilon_{\alpha'}}
\Gamma_{K,J}^{m-}(ii',\sigma \sigma')
\Big|^2
\nonumber \\
&\times
f_F(\epsilon - \mu_\alpha)
\left( 1 - f_F(\epsilon + \Delta_{JK} - \mu_{\alpha'}) \right).
\label{eq:rates_cot}
\end{align}

where the definition of $\Gamma_{K,J}^{m+}(ii',\sigma \sigma') = \langle \Psi_{K}| d_{i \sigma}|\Psi_{m+}^{N+1} \rangle \langle \Psi_{m+}^{N+1} | d^{\dagger}_{i' \sigma'} | \Psi_{J} \rangle$ and $\Gamma_{K,J}^{m-}(ii',\sigma \sigma') = \langle \Psi_{K}| d^{\dagger}_{i \sigma}|\Psi_{m-}^{N-1} \rangle \langle \Psi_{m-}^{N-1} | d_{i' \sigma'} | \Psi_{J} \rangle$, as well as $\Delta_{KJ} = E_K - E_J$ have been introduced. Variables $E_0$ and $E_{m\pm}$ are the energy of the ground state in neutral (N) state  and energies of particular state in the transient charged ($N\pm1$) state, respectively.
Additionally, we explicitly include the Fermi functions ($f_F$), density of states $\rho_\alpha$ and chemical potential of the electrodes $\mu_\alpha$. If we suppose that both Tip and Substrate are made of the same metal in wide-band limit, this is nothing but the Fermi energy of the metal. However, when a bias voltage is applied, the difference between chemical potentials become $\Delta \mu = e \Delta V$. Thus, we will consider a fixed chemical potential for the surface, $\mu_S = E_F$ and we will vary the voltage in the tip electrode, $\mu_T = E_F + e  V$. Note that terms $\Gamma_{K,J}^{m+}(ii',\sigma \sigma')$ and $\Gamma_{K,J}^{m-}(ii',\sigma \sigma')$ represent the simultaneous annihilation and creation of electrons with spin $\sigma, \sigma'$ at different molecular orbital sites, using the charged states ($N\pm1$)  of the quantum dot as intermediates. Hence, they are responsible for spin transitions inside of the dot. Notice also that only $E_0$ appears in the denominators, while the sums run over every charged state $E_{m\pm}$. In doing so, we assume that the ionization energies are much further in energy than the excitations of the neutral system, as pointed out by Delgado and Rossier \cite{delgado2011cotunneling}.

The populations $P_K(V)$ can be computed from the steady-state solution of the rate equations, but in practical applications for small bias voltages it suffices to take the thermal populations.

\section{Origin of asymmetries: dimer model}

Next, we will discuss how low-lying spin excitations can yield asymmetric features in $dI/dV$  spectra. We start by exploring a half-filled extended Hubbard dimer, i.e, a model taking into account HOMO and LUMO orbitals. This model arises naturally from a CAS(2,2) in a lattice model with some local repulsion interaction \cite{ortiz2019exchange}. The possible solutions for such a system are the singlet ($S = 0$, antiferromagnetic coupling) and the triplet ($S = 1$, ferromagnetic coupling) ground state. The solutions are determined by the energetic alignment of the orbitals ($\epsilon_1, \epsilon_2$) and their Coulomb interactions (\emph{U}). To tune the ground state and the excitation singlet-triplet gap for desired values, we introduce \emph{ad-hoc} a term $\delta S^2$ so that the spin excitation value will remain the same for every case. The Hamiltonian $H_{dot} $ reads as follows: 
\begin{align}
H_{\text{dot}} &= \epsilon_1 \left( n_{1\uparrow} + n_{1\downarrow} \right)
+ \epsilon_2 \left( n_{2\uparrow} + n_{2\downarrow} \right)
\nonumber \\
&\quad + U \sum_{i,j} n_{i\uparrow} n_{j\downarrow}
+ U \sum_{i,j} d^{\dagger}_{i \uparrow} d_{j \uparrow}
d^{\dagger}_{j \downarrow} d_{i \downarrow}
\nonumber \\
&\quad + \delta S^2,
\label{eq: homogeneous hubbard dimer}
\end{align}

with $n_{i \sigma} = d_{i\sigma}^\dagger d_{i\sigma}$. 

\begin{figure}[h!]
    \centering
    \includegraphics[width=0.8\linewidth]{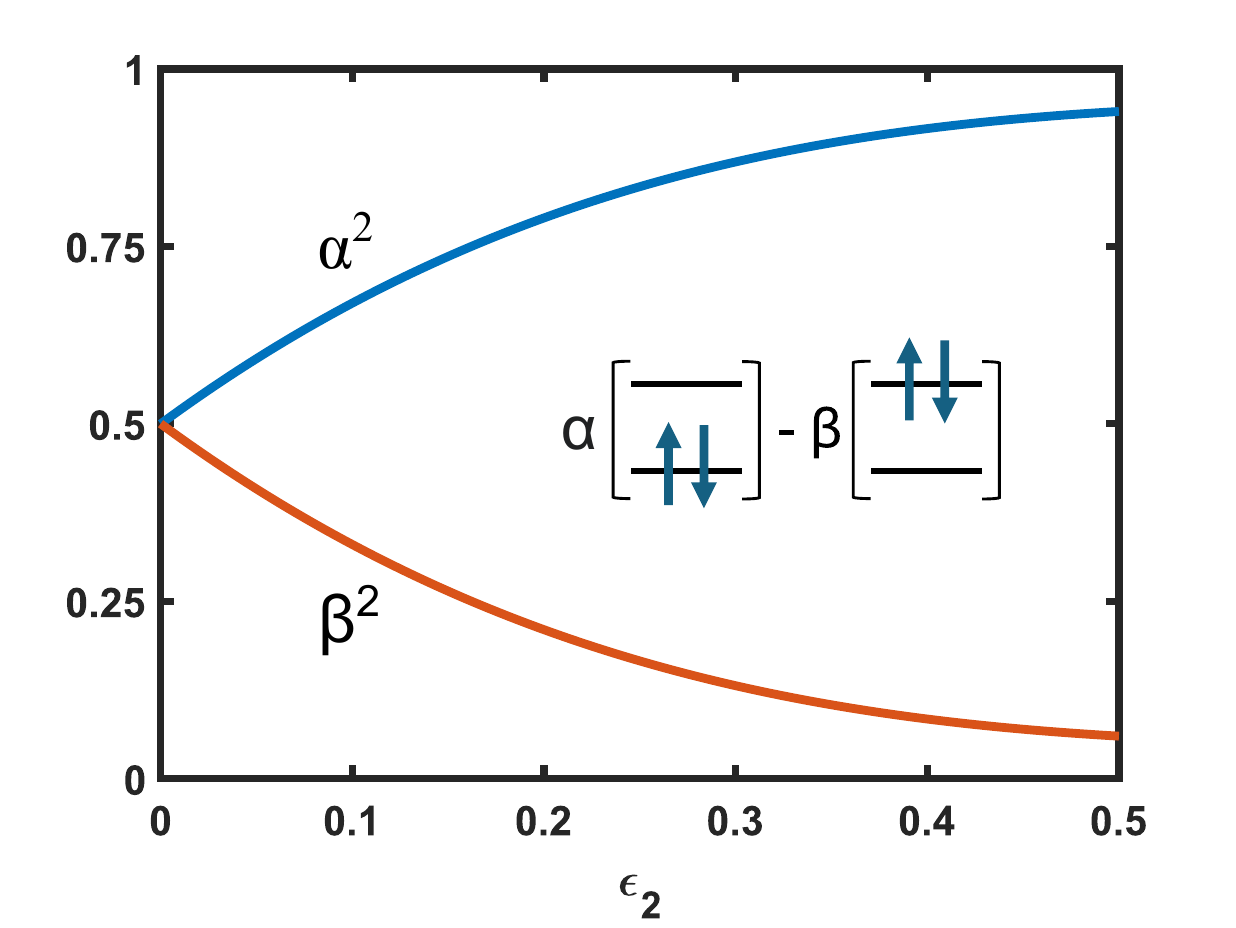}
    \caption{Representation of the singlet state a as function of the energy level $\epsilon_2$ (LUMO) with respect to $\epsilon_1$ (HOMO) and its coefficients $\alpha, \beta$ in the expansion of configurations for $U=230$ meWhen $\epsilon_2 = \epsilon_1$, $\alpha^2 = \beta^2 =0.5$, and we have a perfect diradical with a superposition of two Slater determinants.ts. In the limit of large $\epsilon_2$, $\alpha^2 \approx 1$, $\beta^2 \approx 0$ and the singlet is \emph{closed-shell}.}
    \label{fig:scheme singlet}
\end{figure}

For further discussion, we set $U = 230$ meV, $\epsilon_1 = 0$ and vary $\epsilon_2$ in the range of 0 to 500 meV (adjusting $\delta$ so that the singlet is the ground state with the triplet located at 4 meV). In this way, as the energy difference between the HOMO and LUMO levels increases, the diradical character starts to quench and approaches a \emph{one-electron} picture in which the HOMO is doubly occupied and the LUMO is empty, see FIG. \ref{fig:scheme singlet}. In order to describe the radicality for two levels, it is enough to account for the occupancy of the lowest unoccupied natural orbital (LUNO) \cite{lowdin_natural}, henceforth $n_{LUNO}$. Therefore, $n_{LUNO}=0$ represents a perfect closed-shell system and $n_{LUNO}=1$ a perfect diradical system. Next, the neutral and charged states arising from  eq. \eqref{eq: homogeneous hubbard dimer} for different $\epsilon_2$ are plugged into eq. \eqref{eq:rates_cot}. For simplicity, we assume a symmetric coupling of the quantum dot to the substrate $V_{S1}=V_{S2}\equiv V_{S}$, and set $V_{S}=\sqrt{2}/2$ in all cases. We are stressing here that we are working with electron-hole symmetry. Hence, the energy difference between the neutral ground state and the $N \pm 1$ sectors is the same.   

\begin{figure}[h!]
    \centering
    \includegraphics[width=1\linewidth]{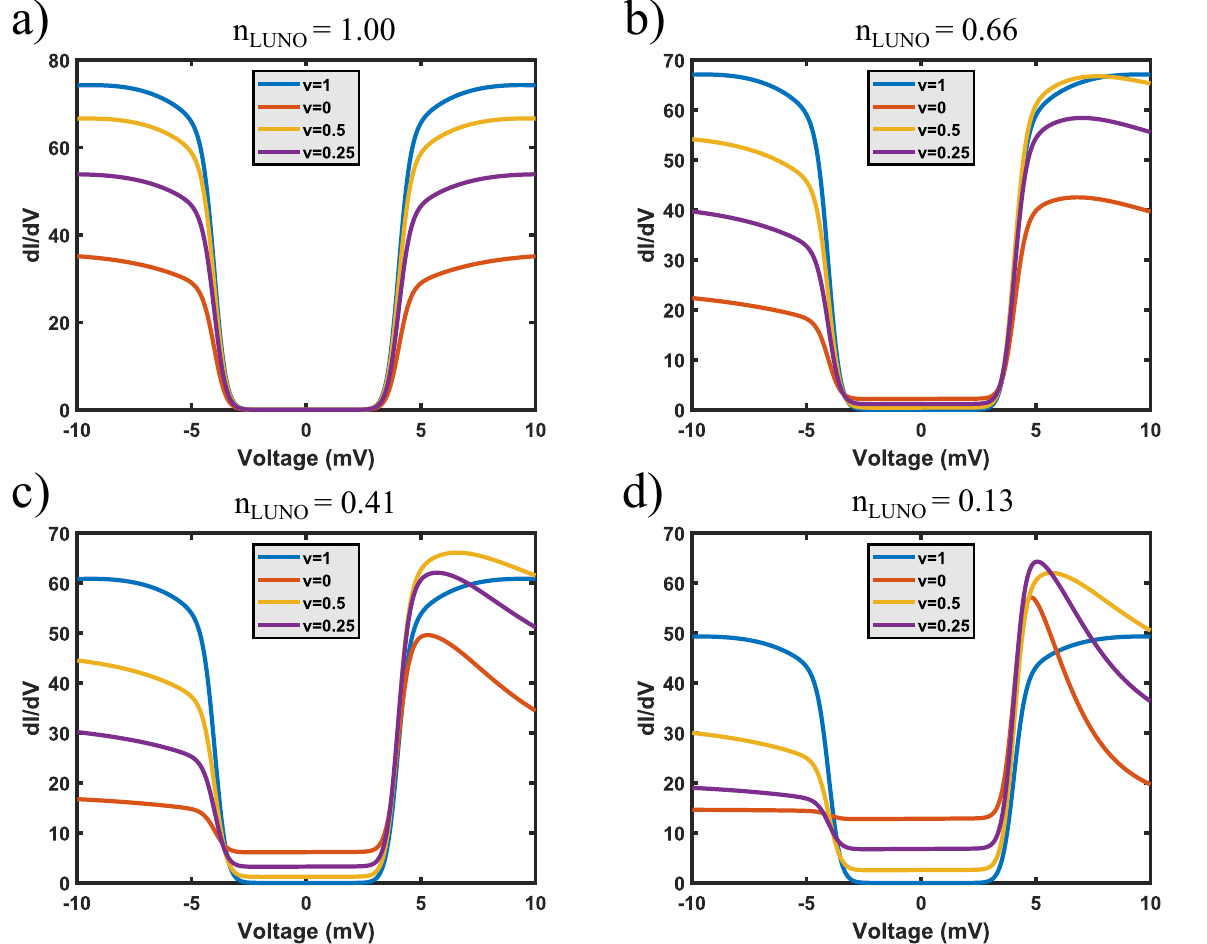}
    \caption{ $dI/dV$  spectra for a two levels extended Hubbard described by the Hamiltonian \eqref{eq: homogeneous hubbard dimer} with a singlet ground state and an excited triplet state at 4 meV for different LUNO occupations. In a-d) the features as function of the relative coupling strength $\text{v}=V_{T1}/V_{T2}$ for $n_{LUNO}=1.00, \, 0.66, \, 0.41, \, 0.13$ respectively.}
    \label{fig: asymmetries}
\end{figure}

\begin{figure}[h!]
    \centering
    \includegraphics[width=1\linewidth]{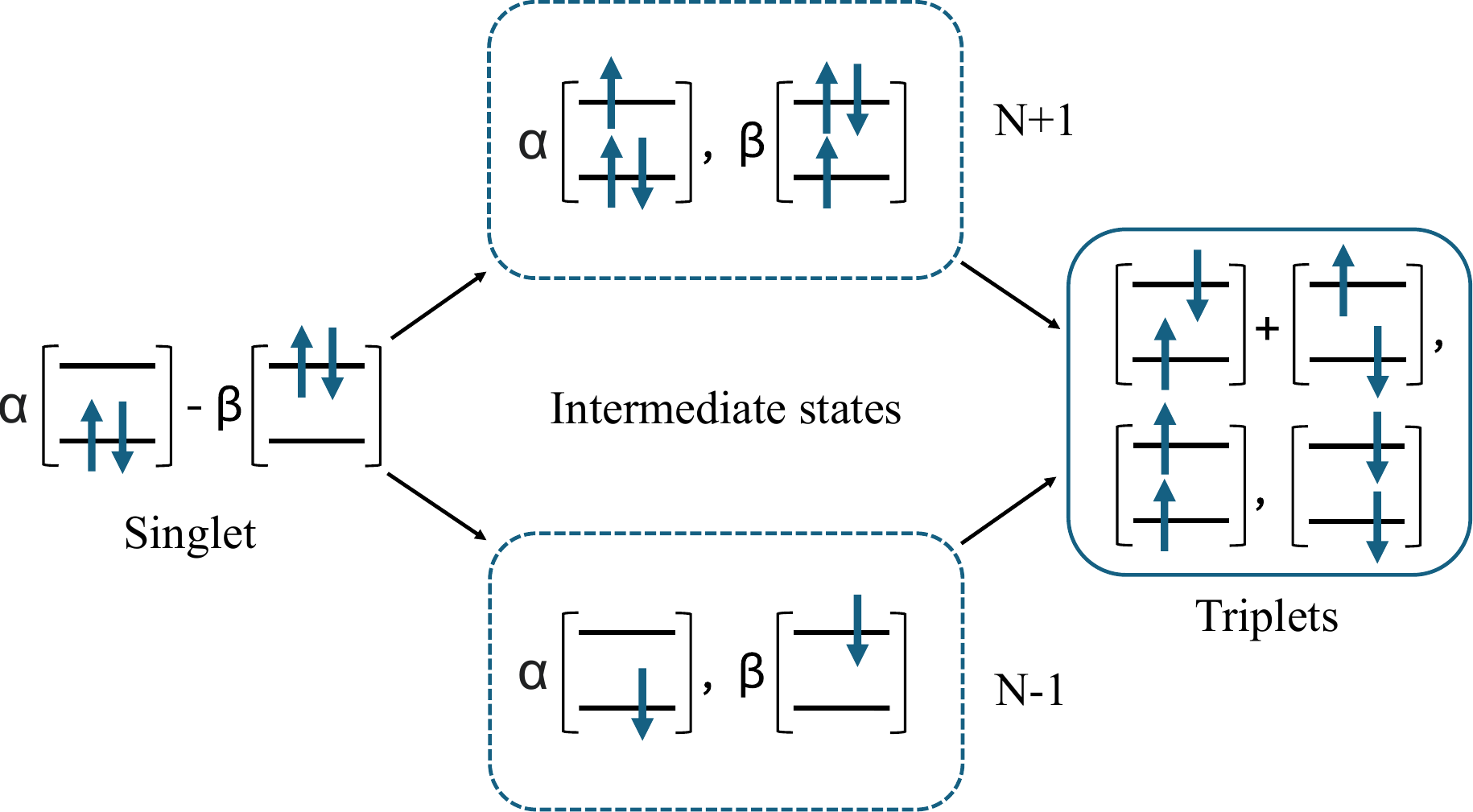}
    \caption{Singlet-triplet excitation through the intermediate charged states as function of the coefficients $\alpha ,\ \beta$ describing the singlet. The initial singlet and coupling to the molecular orbitals filtrates (or preponderates in intensity) the charged and discharged states that then go to the degenerate triplets. For simplicity only the addition and removal of $S_z=1/2$ is plotted, the other spin is completely analogous.} 
    \label{fig:singlet-triplet}
\end{figure}

First, we explore how the asymmetric coupling to the tip affects the line shape of the spin excitation in $dI/dV$  spectroscopy for dimers with different radical character. We consider $V_{T1}^2 + V_{T2}^2  = 1$, and we vary the ratio $\text{v} := V_{T2}/V_{T1}$. In principle, coupling v with different strengths to the two orbital levels is equivalent to moving the STM tip over a molecule with two spatially delocalized radicals. FIG. \ref{fig: asymmetries} displays the simulated $dI/dV$  spectra using eq. \eqref{eq:rates_cot} for different coupling ratios v at different $n_{LUNO}$. When the system is a perfect diradical, we obtain a symmetric line shape with a width equal to twice the excitation energy regardless of the tip coupling ratios, see FIG. \ref{fig: asymmetries} a). As the system starts to quench radicality, the signal becomes more asymmetric when the coupling changes over the two levels until the feature is only observed at positive bias (completely asymmetric) for closed-shell and $\text{v}=0$ (FIG. \ref{fig: asymmetries} d)). Consequently, a tip scan over a molecule with two spatially separated radicals would reveal different profiles for the spin-excitation signals, depending also on the radical character. If the radicality is partially quenched, the $dI/dV$  displays very different signatures from the commonly observed symmetric wells recorded in STS spectroscopy. In general, fully symmetric features can only arise in the case of full radicality (see FIG. \ref{fig: asymmetries}a). \cite{Heinrich2004}. In order to understand the mechanism driving this asymmetric signal, we display the cotunneling scheme of the singlet-triplet excitation in FIG. \ref{fig:singlet-triplet}. Now it is straightforward to see that if only the coupling with the HOMO orbital ($V_{T1}$) is nonzero, the accessibility to the charged states depends on the values of $\alpha ,\ \beta$. Thus, in the case where $\alpha \approx 1 ,\ \beta \approx 0$ the connection to the $N+1$ state is impossible, as it would require the addition of an electron in the LUMO, which would be unfeasible if $V_{T2} = 0$. However, $N-1$ is accessible by directly removing an electron from the HOMO. This explains the characteristic shown in Fig. \ref{fig: asymmetries}  when $\text{v}=0$. For negative voltages, the charging channel is suppressed and no signal is recorded, while for positive voltages, the discharging channel gives a contribution to the signal. When $\alpha \approx \beta$, even for completely asymmetric coupling, both channels have the same intensity. Finally, for intermediate situations, the ratio between $\alpha$ and $\beta$ will tune the intensity of the $N \pm 1$ channels. It is worth noticing that this selection only happens because of the different couplings to the molecular orbital sites whose Slater determinants span the many-body states. Again, it is worth stressing that the asymmetric line shape arises exclusively from the combination of radical quenching and asymmetric coupling, since the system is always at electron–hole symmetry. 

Next, we study the $dI/dV$  profile for the triplet ground state. To achieve this, it is sufficient to further decrease the parameter $\delta$ in \eqref{eq: homogeneous hubbard dimer}. It is important to note that in the case of high-spin ground states, low lying excitations can also arise from magnetic anisotropy effects \cite{Heinrich2004, ternes2015}. In this situation, the lowest allowed transitions do not occur between states of different total spin $S^2$ (multiplet transition), but between different triplet states with different $S_z$ belonging to a multiplet labeled with a certain $S^2 \neq 0$. To account for this effect we introduce a magnetic anisotropy term, $H_{\text{dot}} \rightarrow H_{\text{dot}} + DS_z^2$, splitting the triplet states into two levels: $S_z=0$ and $S_z=\pm1$. 

Now, we explore and compare four different cases: the ferromagnetic dimer (FMD) with the singlet excited state at 4 meV, both without and with magnetic anisotropy ($D = 2$ meV) for $n_{LUNO} = 0.52$ in the singlet; and the FMD without and with magnetic anisotropy ($D = 2$ meV) for $n_{LUNO} = 0.13$ in the singlet. Again, we observe that the symmetric features are only possible when the singlet state is fully diradical.

\begin{figure}[h!]
    \centering
    \includegraphics[width=1.0\linewidth]{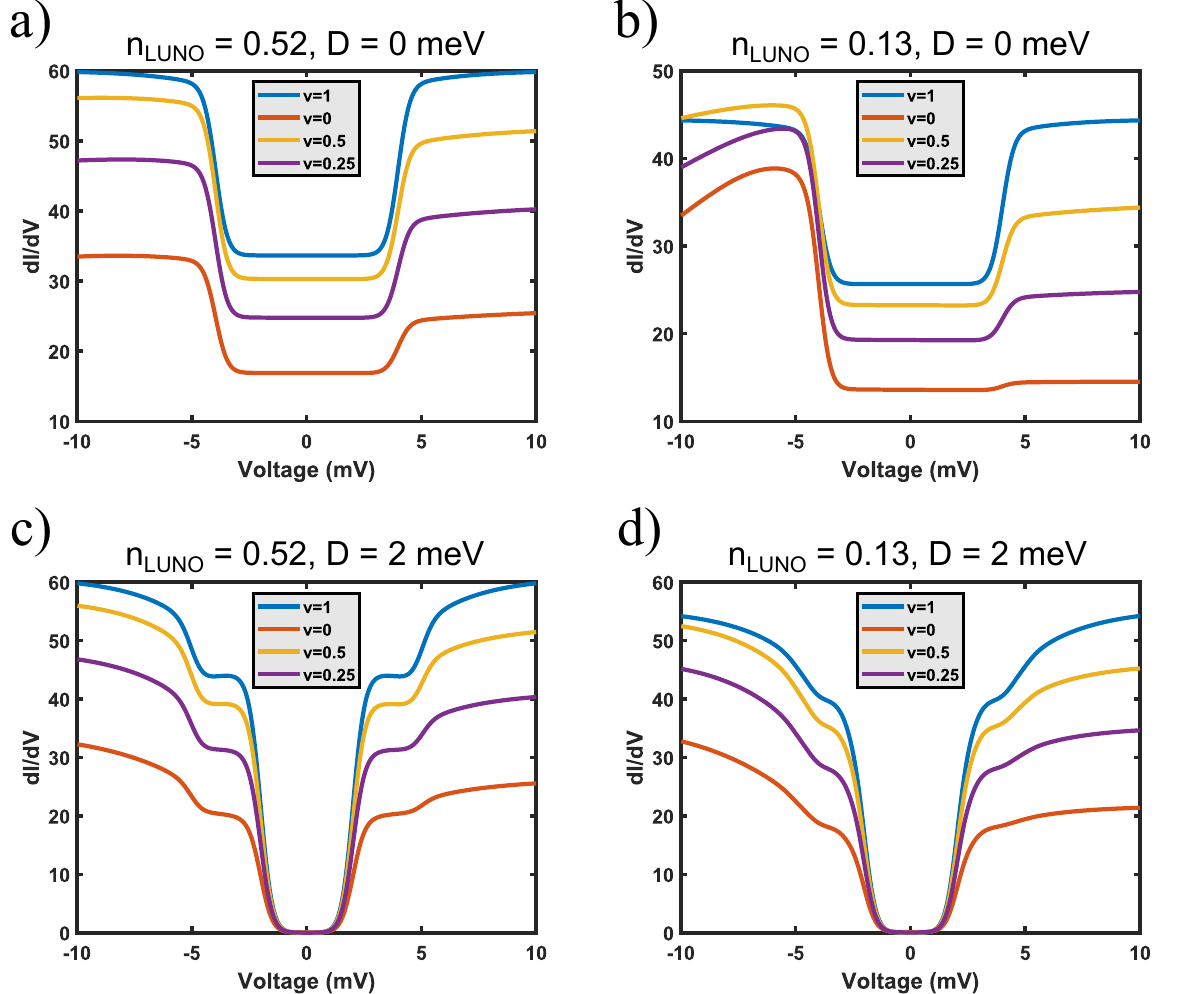}
    \caption{dI/dV spectroscopy for the triplet ground states with singlet situated at 4 meV. In a,b) $\epsilon_2 = 200$ meV and $\epsilon_2 = 500$ meV, so that $n_{LUNO}=0.52$ and $n_{LUNO}=0.13$ in the singlet excited states, respectively. The transitions happen from the degenerated triplets to the singlet, developing an asymmetric character similar to the ones in Fig. \ref{fig: asymmetries}. In c,d) an anisotropy of $D=2$ meV is introduced, splitting the triplets. The transitions occur now between the $S_z=0$ and $S_z= \pm1$ states inside of the triplet. Under this situation no asymmetries are observed.  
    }
    \label{fig: magnetic anisotropy}
\end{figure}

In FIG. \ref{fig: magnetic anisotropy} a,b), the $dI/dV$  profile of the triplet–singlet excitation exhibits asymmetries with respect to zero bias voltage as a function of the tip–coupling ratios. This behavior is completely analogous to the antiferromagnetic case discussed above and becomes more pronounced as the radical character is quenched. Indeed, the transition scheme is the reverse of that presented in FIG. \ref{fig:singlet-triplet}, and consequently it is the opposite branch in the $dI/dV$  signal that vanishes.

When the magnetic anisotropy is switched on at an energy scale comparable to the ferromagnetic coupling, the $dI/dV$  profile becomes symmetric, recovering two visible “arm features” at positive and negative voltages even at the ratio $\text{v} = 0$, independently of the radical character of the singlet (FIG. \ref{fig: magnetic anisotropy} c,d)). The transitions observed in the $dI/dV$  spectra occur now purely between different $S_z$ states within the triplet. Naturally, these transitions take place at the energy splitting value $D$ and are symmetric and independent of the coupling ratios. This is due to the expansion of the triplet states in terms of the molecular orbitals (right part of FIG. \ref{fig:singlet-triplet}). Hence, starting from the $S_z = 0$ triplet, any intermediate charge state is accessible and equally weighted, regardless of charging or discharging, and from there the transition to $S_z = \pm 1$ occurs.

This clearly shows that, in the absence of multiplet excitations, magnetic anisotropy produces symmetric features for any tip position. Moreover, the appearance of strongly asymmetric steps in the $dI/dV$  measurements indicates the presence of low-lying multiplet excitations, whose energies are typically not well captured by one-electron approaches. In addition, we observe that the radical character of the system directly affects the $dI/dV$  line shape for spatially separated radicals. This provides a route to quantify the degree of radical character by shifting the tip position and recording the resulting asymmetry.

\section{Origin of asymmetries: the case of a metal porphyrin }

In this section, we apply the multireference cotunneling framework to a realistic, strongly correlated molecule physisorbed on a metallic surface and analyze the resulting STS line shape.

Metal porphyrins are prototypical correlated molecules that have been widely studied by STM \cite{Sun2020,Zhang2023,Otsuki2010}, and their low-bias STS spectra often display pronounced asymmetries \cite{RubioVerd2018}. Here we focus on cobalt cyano-porphyrin (Co-Por), which exhibits a high-spin multireference electronic structure. The system is related to zinc cyano-porphyrin studied previously \cite{Tenorio2024}, where two unpaired $\pi$ electrons on the ligand form an open-shell singlet ground state. In Co-Por, in addition to the ligand $\pi$ radicals, the Co center contributes unpaired electrons in $d$-like orbitals. Co-Por therefore belongs to the class of high-spin $\pi$--$d$ molecular magnets with strongly correlated electronic structure, which have attracted increasing interest due to magnetic interactions that can go beyond standard superexchange mechanisms \cite{Miyazaki2003,Robles2025,Zhang2025, Sun2022,Frezza2025}.

\begin{figure}[h!]
    \centering
    \includegraphics[width=\linewidth]{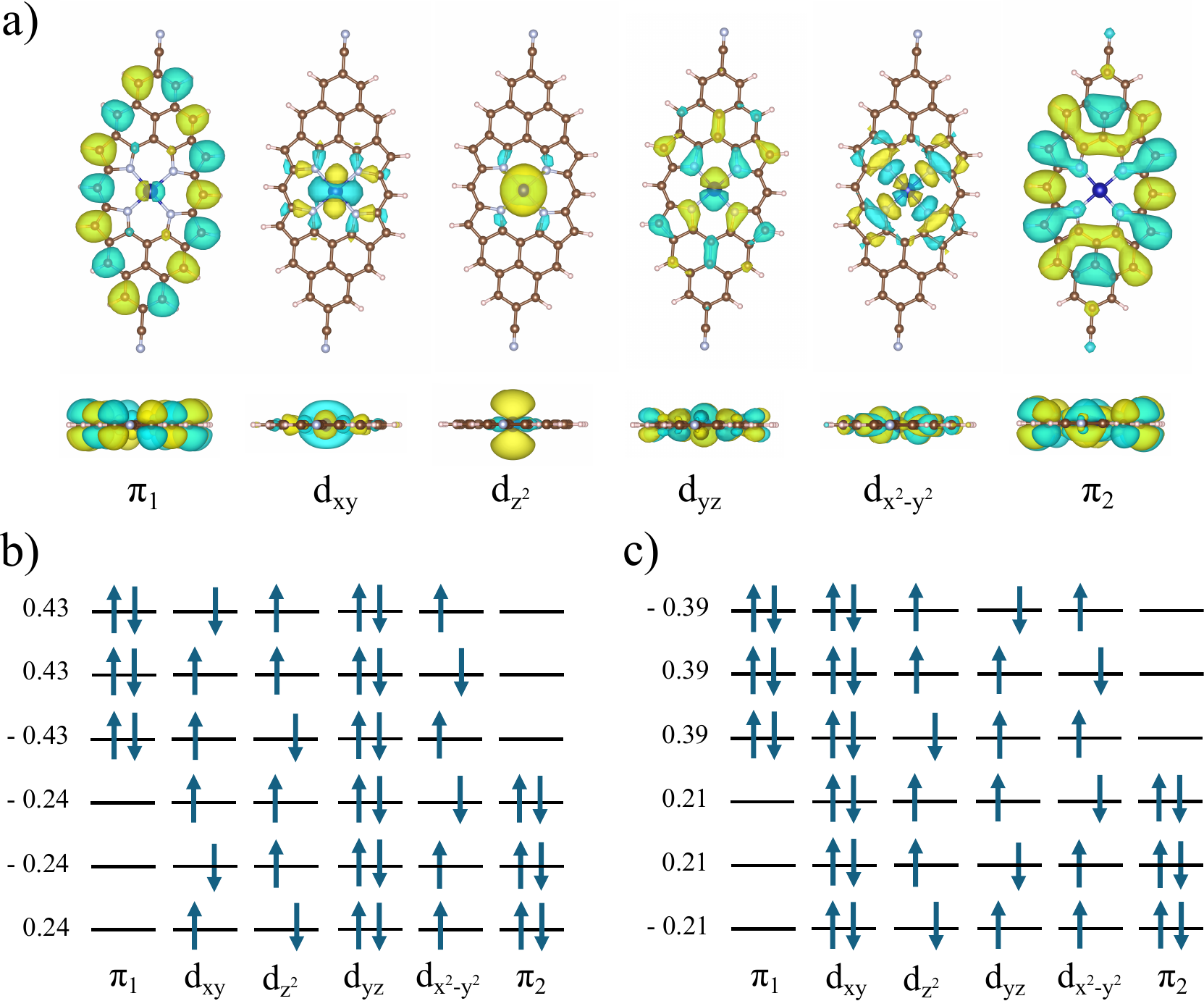}
    \caption{Description of the multireference wavefunction of Co-Por in reduced active space of six molecular orbitals shown in a). Molecular orbitals are displayed from top and side view to reveal their different spatial localization. This six orbitals host 5 unpaired electrons of Co-Por, with three unpaired electrons localized in $d$-like orbitals of cobalt center and other two in the $\pi$-orbitals on the ligand. Lower panel shows the expansion in Slater determinants in the reduced active space for the quartet ground state (b) and the first excited quartet state (c).}
    \label{fig:natura_orbitals}
\end{figure}

First, we carried out \emph{ab initio} CASCI calculations up to the well-converged CASCI(13,13) to analyze in details the electronic structure of Co-Por. We assume that the interaction between the molecule and surface is driven by dispersion interaction and do not alter significantly the electronic structure of the Co-Por. The employed structure results from a geometrical optimization of Co-Por on Au(111) using total energy DFT calculations with  PBE0 exchange–correlation functional \cite{perdew1996rationale}.  
\\ 

The CASCI calculation yields a rich spectrum of many-body eigenstates. The ground state is a quartet ($S=3/2$) state with the occupancy of natural orbitals revealing the presence of five unpaired electrons. The lowest excited states consist of another quartet and a sextet state at 26 and 102 meV above the ground state, respectively. We should note that these energy values are most likely overestimated, as the multireference calculation does not take into account the possible screening effects due to metallic substrate \cite{Zuzak2024} nor the dynamical correlation \cite{guo2017explicitly}.

Unfortunately, calculating the tunneling current using eqs. \eqref{eq:rates_cot} and \eqref{eq: current cotunneling} with the large active space of thirteen orbitals are computationally intractable. However, analysis of the expansion of the Slater determinants describing the ground and low-energy excited states shows that several molecular orbitals in the active space are either fully occupied or completely empty with no influence on the character of the ground and excited states. Consequently, for our purposes, we reduce the molecular orbitals in the active space to six important MOs, see FIG. \ref{fig:natura_orbitals} a). Such a reduced active space is still able to provide the correct character of the ground (quartet) and low-energy excited (quartet, sextet) states, as well as the presence of five unpaired electrons in these orbitals for the ground and first excited states. Interestingly, these five unpaired electrons remain spatially separated, two $\pi$ electrons delocalized over the porphyrin ligand and the other three located in \emph{d} orbitals of Co center. In FIG. \ref{fig:natura_orbitals} b), the three electrons located at  Co center occupy  $d_{xy}$, $d_{z^2}$, and $d_{x^2-y^2}$-like orbitals in the ground state. For the first excited state, these electrons are localized in the $d_{z^2}$, $d_{yz}$, and $d_{x^2-y^2}$-like orbitals, as shown in FIG. \ref{fig:natura_orbitals} c). Hence, the first excitation occurs exclusively within the active space of $d$-like orbitals of the Co center, corresponding to an electron transfer from $d_{xy}$ to $d_{yz}$, while the occupancy of $\pi$-orbitals remains unchanged.  

Because of this spatial separation of the radicals and the excitation, we further partition the system and focus on the metal center. To this end, we retain the four \emph{d} orbitals shown in FIG. \ref{fig:natura_orbitals} with five electrons. We then include the interactions with the remaining occupied orbitals at the mean-field level, thereby constructing an effective tetramer Hamiltonian that describes excitations occurring solely at the metal center. 

To obtain energy scales typical of spin excitations, we reintroduce the term $\delta S^2$ of eq. \eqref{eq: homogeneous hubbard dimer} and tune it to adjust the spectral gap to 4 meV. Notice that this does not affect the electronic configuration of each state, but just their relative energy gaps. The lower spin of the excited state (quartet in the full porphyrin and doublet in the metallic subsystem) is explained by a change in the $\pi$ subsystem. As can be seen by comparing the coefficients in FIG. \ref{fig:natura_orbitals} b,c), the $\pi$ system is in a singlet configuration in the former case and in a triplet configuration in the latter. This simplified tetramer model can now be inserted into eq. \eqref{eq: basic model} as $H_{\text{dot}}$ and solved in the cotunneling regime.

The key feature of the orbitals forming the basis of this tetramer is their markedly different spatial distributions, particularly along the \emph{z} direction (see the side view of the orbitals in FIG. \ref{fig:natura_orbitals} a)). As a consequence, the tip coupling to these molecular levels is expected to be highly asymmetric for a given tip position above the molecule, even though they occupy the same region of the molecule.

\begin{figure}[h!]
    \centering
    \includegraphics[width=0.95\linewidth]{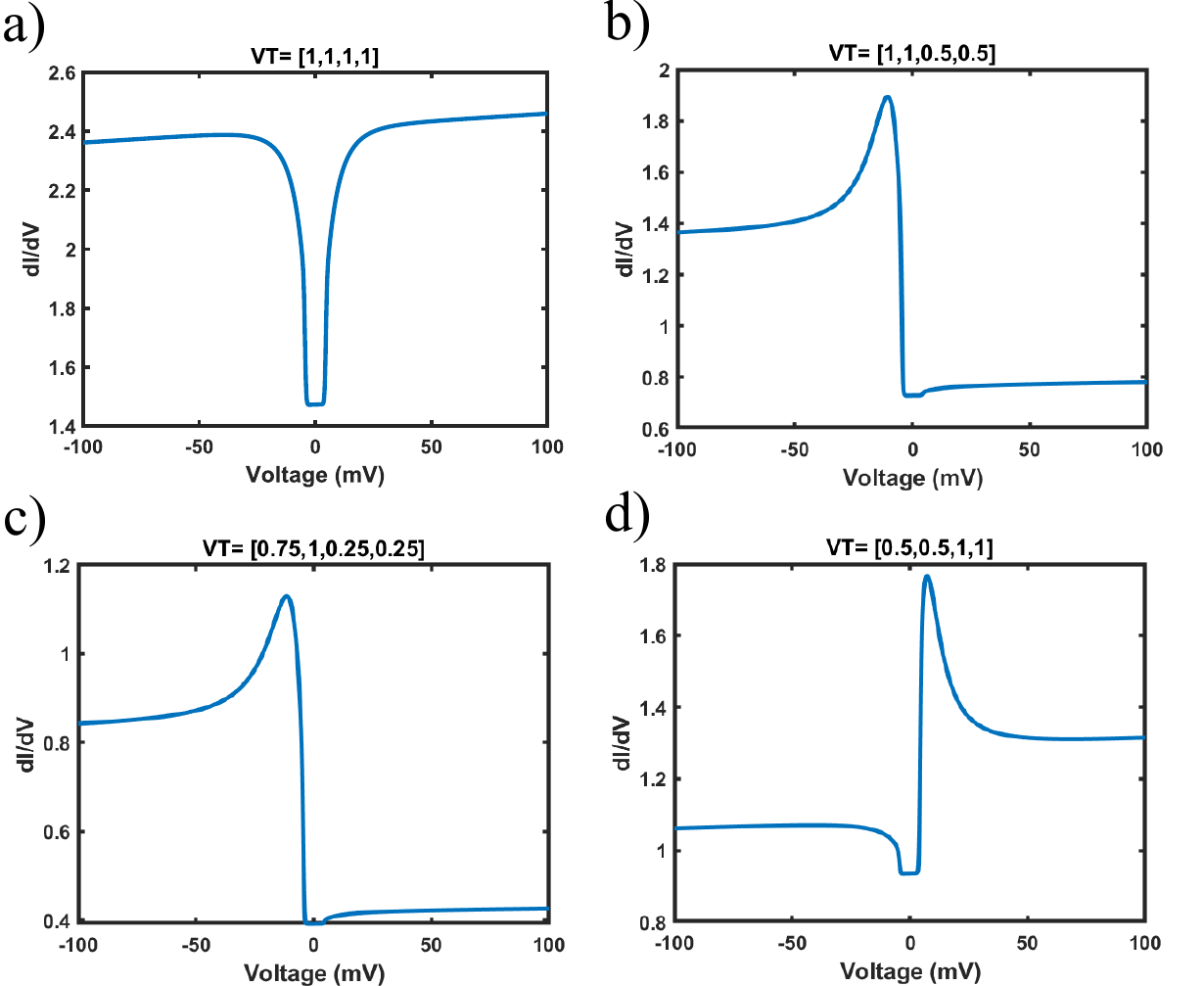}
    \caption{$dI/dV$ spectra simulated for the sub-tetramer Hamiltonian in Co-porphyrin, VT states for tip couplings with $[d_{xy}, d_{z^2}, d_{yz}, d_{x^2 - y^2}]$ orbitals in Fig. \ref{fig:natura_orbitals}. Coupling with surface is kept constant and equal for the three orbitals  (VS$= [1.2, 1.2, 1.2, 1.2]$) and the one with the tip is changed.}
    \label{fig:porphyrin}
\end{figure}

In FIG. \ref{fig:porphyrin}, we show the $dI/dV$  results obtained from such a tetramer model with different tip couplings, defined as $V_T = [d_{xy}, d_{z^2}, d_{yz}, d_{x^2 - y^2}]$. In panel a), homogeneous couplings are applied, and the typical feature of a spin excitation appears as a symmetric dip.

By inspecting the spatial distribution of the orbitals along the \emph{z} direction (Fig. \ref{fig:natura_orbitals}a)), we infer that the tip coupling should be maximal for the $d_{z^2}$ orbital, together with $d_{xy}$. Accordingly, we set $V_{d_{z^2}} = V_{d_{xy}} = 1$ and decrease the remaining couplings according to their \emph{z}-direction distributions. In FIG. \ref{fig:porphyrin} b,c), we show the resulting $dI/dV$  profiles for $V_T = [1, 1, 0.5, 0.5]$ and $V_T = [0.75, 1, 0.25, 0.25]$. In both cases, a strongly asymmetric signal is obtained, with a peak located at negative bias and a vanishing signal at positive bias. 

We emphasize that, in all of these profiles, the molecular electronic structure of neutral ($N$) and charged ($N\pm1$) states is fixed.  The only ingredient providing asymmetric STS line shape is the asymmetric coupling between the electronic states of the tip and one-electron molecular orbitals, described by the tunneling Hamiltonian, $V_T$. Precisely, those one-electron molecular orbitals with varying electron configuration in the expansion of many-body wavefunction of the ground-state and first-excited-state wavefunctions contribute mostly to the origin of the asymmetry of the STS line shape. For example, Fig  \ref{fig:natura_orbitals} b,c) displays relevant electronic configurations contributing to the ground and first excited states of Co-Por, in which several $d$-like molecular orbitals change their electronic configuration ( e.g $d_{xy},d_{z^2},d_{x^2-y^2}$).

Fig. \ref{fig:porphyrin} d) displays resulting STS linehapes with the modified couplings $V_T = [0.5, 0.5, 1, 1]$. Although this configuration is unrealistic given the spatial extent of the \emph{d} orbitals, it illustrates how the negative-bias peak disappears while the positive-bias peak remains. This behavior indicates that the mediated charged states $N \pm 1$ correspond to adding electrons to $d_{xy}$ and $d_{z^2}$ orbitals and removing electrons from $d_{yz}$ and $d_{x^2-y^2}$ orbitals, leading finally to an excited state with reduced radicality. In this sense, the transition scheme is completely analogous to that described in Fig. \ref{fig:singlet-triplet}, though involving more complex state expansions.

\section{Conclusions}
We presented the multireference cotunneling theory of molecular quantum dots weakly coupled to metallic surfaces.
Our findings shed new light on the microscopic origin of asymmetric inelastic conductance features, which are often interpreted exclusively in terms of Kondo physics or spin–orbit–induced anisotropy. In particular, we show that for a given spin excitation between molecular states with different total spin, the unequal coupling strengths of the STM tip with the different one-electron molecular orbitals with varying electron configuration forming a basis of multiplets in many-body wavefucntion leads to sharply asymmetric $dI/dV$ line shapes.  Our results provide a new framework for interpreting low-energy STS line shapes beyond particle-hole symmetry and broaden the understanding of IETS spectroscopy in molecular systems.

\section{Acknowledgment} 
We acknowledge the support from GACR 25-17866X and the CzechNanoLab Research Infrastructure supported by MEYS CR (LM2023051). We also acknowledge the financial support from the TERAFIT project (CZ.02.01.01/00/22\_008/0004594). 

\bibliography{bibliography.bib}

\end{document}